# A poroelastic model coupled to a fluid network with applications in lung modelling

Lorenz Berger*, Rafel Bordas†, Kelly Burrowes‡,
Vicente Grau§and David Kay ¶
Department of Computer Science, University of Oxford, Oxford, UK
AND
SIMON TAVENER‖
Department of Mathematics, Colorado State University, Fort Collins, USA

April 19, 2015

**Abstract**

Here we develop a lung ventilation model, based a continuum poroelastic representation of lung parenchyma and a 0D airway tree flow model. For the poroelastic approximation we design and implement a lowest order stabilised finite element method. This component is strongly coupled to the 0D airway tree model. The framework is applied to a realistic lung anatomical model derived from computed tomography data and an artificially generated airway tree to model the conducting airway region. Numerical simulations produce physiologically realistic solutions, and demonstrate the effect of airway constriction and reduced tissue elasticity on ventilation, tissue stress and alveolar pressure distribution. The key advantage of the model is the ability to provide insight into the mutual dependence between ventilation and deformation. This is essential when studying lung diseases, such as chronic obstructive pulmonary disease and pulmonary fibrosis. Thus the model can be used to form a better understanding of integrated lung mechanics in both the healthy and diseased states.

## 1 Introduction

The main function of the lungs is to exchange gas between air and blood, supplying oxygen during inspiration and removing carbon dioxide by subsequent expiration. Gas exchange is optimised by ensuring efficient matching between ventilation and blood flow, the distributions of which are largely governed by branching structure of the airway and vascular trees, tissue deformation and gravity. In this work, we focus on the link between tissue deformation and ventilation. Previous work of whole organ lung models has typically focused on modelling either ventilation or tissue deformation [1, 2, 3, 4, 5, 6, 7, 8]. However, the two components are intrinsically linked; evaluating them separately does not necessarily provide an accurate description of lung function. To gain a better understanding of the biomechanics in the lung it is therefore necessary to fully couple the tissue deformation with the ventilation. To achieve this tight coupling between tissue deformation and ventilation we propose a novel model that approximates the lung parenchyma by a biphasic (tissue and air) poroelastic model, that is coupled to an airway network model. A poroelastic model of lung parenchyma was first proposed in [9]. This model was applied in a very simple 2D geometrical domain. In [10] a homogenisation approach was taken to derive macroscopic poroelastic equations for average air flows and tissue displacements in lung parenchyma under high frequency ventilation.

*Email: lorenz.berger@cs.ox.ac.uk
†Email: rafel.bordas@cs.ox.ac.uk
‡Email: kelly.burrowes@cs.ox.ac.uk
§Email: vicente.grau@oerc.ox.ac.uk
¶Email: david.kay@cs.ox.ac.uk
‖Email: tavener@math.colostate.edu

The fully coupled poroelastic fluid-network model developed here has the potential to advance the understanding of the biomechanics of respiratory diseases. It takes advantage of realistic deformation boundary conditions and incorporates several generations of a realistic airway tree, both of which are obtained from imaging data. It has the ability to define localized material properties. An integrated model of ventilation and tissue mechanics is particularly important for understanding respiratory diseases since nearly all pulmonary diseases lead to some abnormality of lung tissue mechanics [11]. Chronic obstructive pulmonary disease (COPD) encompasses emphysema, the destruction of alveolar tissue, as well as chronic bronchitis, which can cause severe airway remodeling, bronchoconstriction and air trapping. All of the above affect tissue deformation since sections of lung are either not able to expand to inspire air, or to contract to release air. The effects of physiological changes occurring during disease, such as airway narrowing and changes in tissue properties, on regional ventilation and tissue stress are not well understood. For example, one hypothesis is that airway disease may precede emphysema [12]. An integrated model of ventilation and tissue mechanics can be used to investigate the impact of airway narrowing and tissue stiffness during obstructive lung diseases on tissue stresses, alveoli pressure and ventilation.

Ventilation models that incorporate many independent elastic alveolar units, e.g., [1, 2] are commonly known as balloon models. In these models, alveolar units can expand and contract independently of the sizes and positions of neighbouring units, which can lead to unphysical solutions in which multiple balloons apparently occupy the same location in space. In reality, acini do not function as independent elastic balloons [13]. They are physically coupled through fibrous scaffolding and shared alveolar septa. In Figure 1 we illustrate this issue by comparing two simplistic balloon models. Our model does not suffer from this difficulty.

Lung models [1, 2] provide information regarding the distribution of flow within the lung as a result of a pleural pressure boundary condition. However it is not possible to experimentally measure the pleural pressure in vivo using imaging or other apparatus. As part of the simulation protocol, the pleural pressure is therefore often tuned until physiological realistic flow rates are achieved. To overcome this issue, [14, 15] proposed to estimate the flow boundary conditions for full organ ventilation models by means of image registration. However, by relying on image registration to determine the ventilation distribution within the tissue, one is not able to model the change in ventilation distribution due to progression of disease. Following the approach in [15], we integrate image registration based boundary conditions within the proposed poroelastic model of lung deformation. In particular, we register expiratory images to the inspiratory images, to yield an estimate of the deformation boundary condition for the lung surface, and drive the simulation through this deformation boundary condition. Thus the tissue deformation and subsequent flow boundary condition for tree branches inside the lung and ventilation distribution is not pre-determined, but calculated from the coupled poroelastic-airway-tree model.

The rest of this paper is organized as follows. In section 2 we present the assumptions inherent in a poroelastic model for the lung. We develop a poroelastic lung model in section 3, discuss its discretization in section 4 and describe the generation of realistic boundary and initial conditions in section 5. We present numerical simulations of tidal breathing in section 6 and investigate the effect of airway constriction and tissue weakening. Finally, we draw some conclusions in section 7.

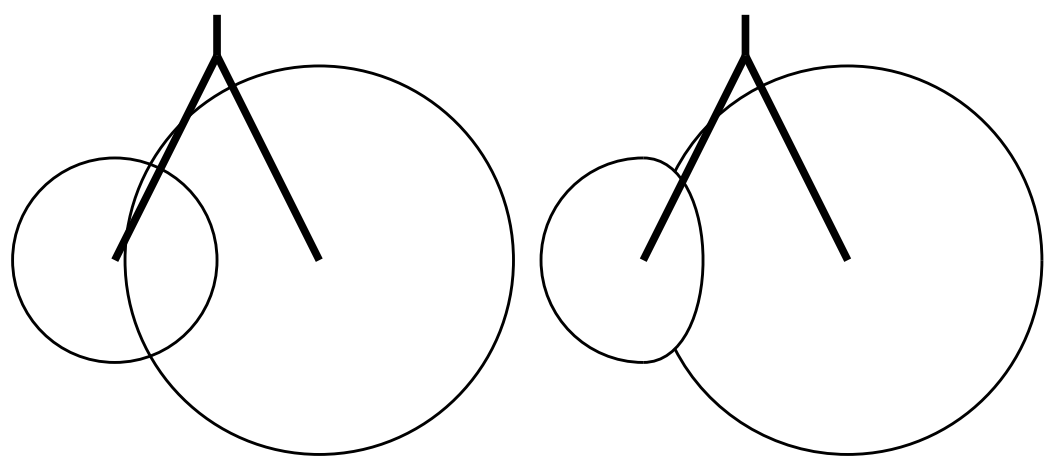

Figure 1: Sketch of two balloon models where the right alveolar unit is more compliant. (a) Balloon model with independent alveolar units. The overlap is physically unrealistic. (b) Balloon model where the alveolar units are coupled. Here the inflation of each alveolar unit is compromised by the expansion of its neighbor.

## 2 Modeling lung parenchyma as a (continuous) poroelastic material

In order to make a computationally tractable lung model, we assume that we can approximate the lung parenchyma as a (continuous) poroelastic materia and couple this to an airway tree network. The use of a continuum model can be supported by consideration of the different length scales and structures of the tissue. For the microscopic length scale, denoted by $l$, we will use the diameter of an alveolus which is approximately 0.02 cm [16]. The macroscopic length scale $L$ can be taken to be the diameter of a lung segment which measures around 4 cm. The ratio of the different length scales is therefore $\epsilon := \frac{l}{L} \approx 0.005 \ll 1$. This, together with the assumption that the structure of an acinus is porous (see Figures 2a and 2b) and approximately periodic, supports the use of averaging techniques over the tissue to obtain a continuum description in the form of a poroelastic medium. By averaging over the tissue we do not seek to model individual alveoli but introduce macroscopic parameters such as the permeability and elasticity coefficients. In general, lung diseases affect significant regions of alveoli (lung tissue), and thus, by changing the macro-scale parameters over the affected area of tissue we are able to model changes in the tissue due to disease.

We make the following assumptions.

**(a)** Blood is simply part of the tissue (solid phase); The volume occupied by blood in the capillaries is similar to the volume occupied by collagen and elastin fibers (illustrated in Figure 2a). The space not occupied by air is about 7% of the parenchymal volume and is made up of 50% capillary blood and 50% of collagen and elastin fibers [17]. Further, the density of blood is similar to the density of tissue and much larger than that of air ($1060\ \mathrm{kg\,m^{-3}} \gg 1.18\ \mathrm{kg\,m^{-3}}$).

**(b)** Both the solid phase (blood and tissue) and a fluid phase (air) are incompressible [1].

**(c)** Accelerations of the solid phase in the poroelastic medium are negligible and will be ignored; Calculations of the sinusoidal motion of tissue near the diaphragm during normal breathing yield an estimate of $0.02\,\mathrm{ms}^{-2}$ for the maximum acceleration of lung parenchyma, which is negligible compared to the acceleration of gravity

**(d)** We ignore accelerations of the fluid phase within the poroelastic medium [18]; The Reynolds number in the lower airways that form part of the lung parenchyma has been estimated to be between 0.01 and 1 [19].

**(e)** We neglect the fluid viscous stress; A dimensional analysis [20] shows that when the microscopic and macroscopic length scales are widely separated, the internal viscous stresses in the fluid are small compared to the drag forces between the fluid and the porous structure.

**(f)** The flow in the airways is laminar (Poiseuille) flow [2, 21]; Diseases affecting the airway tree will be modelled effectively by changing resistance (airway radius) parameters in the network flow model.

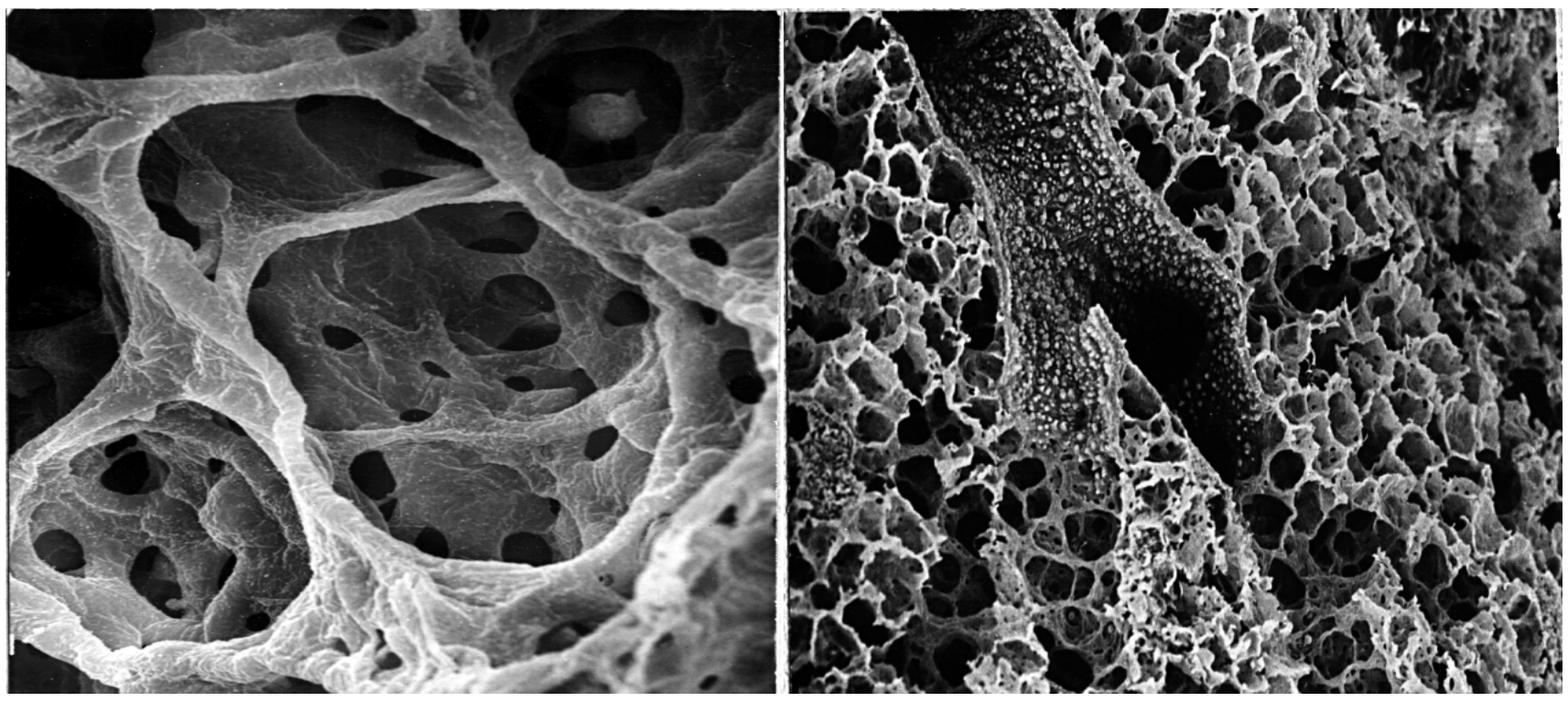

(a) (b)

Figure 2: (a) Alveoli in an alveolar duct. The dark round openings are pores between alveoli. The alveolar wall is quite thin and contains a network of capillaries. The average diameter of one alveoli is 0.2 mm. (b) Transition from terminal bronchiole to alveolar duct, from conducting airway to oxygen transfer area, diameter of terminal bronchiole is 0.5 mm. Images are reproduced from [22].

# 3 The poroelastic lung model

We give a short review of the development of a model for a poroelastic medium undergoing large deformations. A more detailed description and full derivation of the poroelastic equations can be found in [23] and [24].

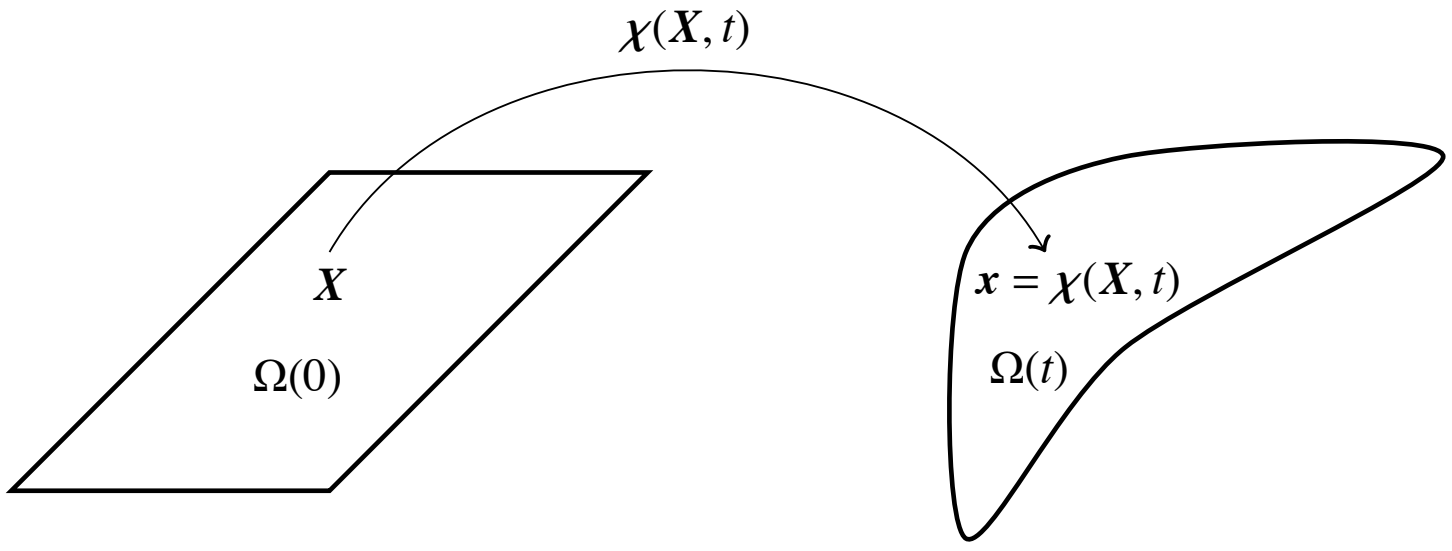

Figure 3: Illustration of the solid deformation.

### 3.1 Kinematics

Let the volume $\Omega(0)$ be the undeformed Lagrangian (material) reference configuration and let $\boldsymbol{X}$ indicate the position of a particle in $\Omega(0)$ at $t = 0$. The position of a particle in the deformed configuration $\Omega(t)$ at time $t > 0$ is given by $\boldsymbol{x}$, with $\boldsymbol{x} = \boldsymbol{\chi}(\boldsymbol{X}, t)$ as shown in Figure 3. The deformation map, $\boldsymbol{\chi}(\boldsymbol{X}, t)$, is a continuously differentiable, invertible mapping from $\Omega(0)$ to $\Omega(t)$. Thus the inverse of the deformation map, $\boldsymbol{\chi}^{-1}(\boldsymbol{x}, t)$, is such that $\boldsymbol{X} = \boldsymbol{\chi}^{-1}(\boldsymbol{x}, t)$. The displacement field is given by

$$\boldsymbol{u}(\boldsymbol{X}, t) = \boldsymbol{\chi}(\boldsymbol{X}, t) - \boldsymbol{X}. \tag{1}$$

The deformation gradient tensor is

$$\boldsymbol{F} = \frac{\partial \boldsymbol{\chi}(\boldsymbol{X}, t)}{\partial \boldsymbol{X}}, \tag{2}$$

and the symmetric right Cauchy-Green deformation tensor is

$$\boldsymbol{C} = \boldsymbol{F}^T \boldsymbol{F}. \tag{3}$$

The Jacobian is defined as

$$J = \det(\boldsymbol{F}), \tag{4}$$

and represents the change in an infinitesimal control volume from the reference to the current configuration, i.e.,

$$d\Omega(t) = J d\Omega(0). \tag{5}$$

Note that $J > 0$.

### 3.2 Volume fractions

We consider saturated porous media in which the fluid accounts for volume fractions $\phi_0(\boldsymbol{X}\ , t\ =\ 0)$ and $\phi(\boldsymbol{x}, t)$ of the total volume in the reference and current configurations respectively, where $\phi$ is known as the porosity. The fractions for the solid are therefore $(1 - \phi_0)$ and $(1 - \phi)$ in the reference and current configuration respectively. For the mixture, $\rho$ is the density in the current configuration given by

$$\rho = \rho^s(1 - \phi) + \rho^f \phi \quad \text{in } \Omega(t), \tag{6}$$

where $\rho^s$ and $\rho^f$ are the densities of the fluid and solid, respectively. We assume that both the solid and the fluid are incompressible so that $\rho^s = \rho_0^s$ and $\rho^f = \rho_0^f$. Although both the solid and fluid are assumed to be incompressible, the control volume can expand or contract due to fluid entering or leaving the region, and

$$J = \frac{1-\phi_0}{1-\phi}. \tag{7}$$

### 3.3 The poroelastic model for lung parenchyma

We define the boundary $\partial\Omega(t) = \Gamma_d(t) \cup \Gamma_n(t)$ for the mixture and $\partial\Omega(t) = \Gamma_p(t) \cup \Gamma_f(t)$ for the fluid, with an outward pointing unit normal $\boldsymbol{n}$. We seek displacement $\boldsymbol{\chi}(\boldsymbol{X}, t)$, fluid flux $\boldsymbol{z}(\boldsymbol{x}, t)$ and pressure $p(\boldsymbol{x}, t)$ such that

$$\left.\begin{aligned}
-\nabla \cdot (\boldsymbol{\sigma}_e - p\boldsymbol{I}) &= \rho \boldsymbol{f} && \text{in } \Omega(t),\\
\boldsymbol{k}^{-1}\boldsymbol{z} + \nabla p &= \rho^f \boldsymbol{f} && \text{in } \Omega(t),\\
\nabla \cdot (\boldsymbol{\chi}_t + \boldsymbol{z}) &= g && \text{in } \Omega(t),\\
\boldsymbol{\chi}(\boldsymbol{X}, t)|_{\boldsymbol{X}=\boldsymbol{\chi}^{-1}(\boldsymbol{x},t)} &= \boldsymbol{X} + \boldsymbol{u}_D && \text{on } \Gamma_d(t),\\
(\boldsymbol{\sigma}_e - p\boldsymbol{I})\boldsymbol{n} &= \boldsymbol{t}_N && \text{on } \Gamma_n(t),\\
\boldsymbol{z} \cdot \boldsymbol{n} &= q_D && \text{on } \Gamma_f(t),\\
p &= p_D && \text{on } \Gamma_p(t),\\
\boldsymbol{\chi}(\boldsymbol{X}, 0) &= \boldsymbol{X} && \text{in } \Omega(0).
\end{aligned}\right\} \tag{8}$$

The fluid flux $\boldsymbol{z} = \phi(\boldsymbol{v}^f - \boldsymbol{v}^s)$ where $\boldsymbol{v}^f$ and $\boldsymbol{v}^s$ are the velocities of the fluid and solid components respectively, and $\boldsymbol{\sigma}_e$ is the stress tensor given by

$$\boldsymbol{\sigma}_e = \frac{1}{J}\boldsymbol{F} \cdot 2\frac{\partial W(\boldsymbol{C})}{\partial \boldsymbol{C}} \cdot \boldsymbol{F}^T, \tag{9}$$

where $W(\boldsymbol{C})$ denotes a strain-energy law (hyperelastic Helmholtz energy functional) dependent on the deformation of the solid. To make the interpretation of the elasticity constants and dynamics of the model as simple as possible we chose a Neo-Hookean law taken from [25], with the penalty term chosen such that $0 \le \phi < 1$,

$$W(\boldsymbol{C}) = \frac{\mu}{2}(\mathrm{tr}(\boldsymbol{C}) - 3) + \frac{\lambda}{4}(J^2 - 1) - (\mu + \frac{\lambda}{2})\ln(J - 1 + \phi_0). \tag{10}$$

The material parameters $\mu$ and $\lambda$ can be related to the more familiar Young's modulus $E$ and the Poisson ratio $\nu$ by $\mu = \dfrac{E}{2(1+\nu)}$ and $\lambda = \dfrac{E\nu}{(1+\nu)(1-2\nu)}$. The permeability tensor is given by

$$\boldsymbol{k} = J^{-1}\boldsymbol{F}\boldsymbol{k}_0(\boldsymbol{C})\boldsymbol{F}^T, \tag{11}$$

where $\boldsymbol{k}_0(\boldsymbol{C})$ is the permeability in the reference configuration, which may be chosen to be some (nonlinear) function dependent on the deformation. Examples of deformation dependent permeability tensors for biological tissues can be found in [26, 27, 28]. We will use the law that has been proposed in [26] to model lung parenchyma,

$$\boldsymbol{k}_0 = k_0 \left(J\frac{\phi}{\phi_0}\right)^{2/3} \boldsymbol{I}. \tag{12}$$

### 3.4 Network flow model for the airway tree

The flow rate $Q_i$ through the $ith$ segment in the airway network is given by the pressure-flow relationship

$$P_{i,1} - P_{i,2} = R_i Q_i, \tag{13}$$

where $R_i = \dfrac{8l\mu_f}{\pi r^4}$ is the Poiseuille flow resistance of a pipe segment, where $r$ and $l$ are the radius and length of the pipe, $\mu_f$ is the dynamic viscosity, and $P_{i,1}$ and $P_{i,2}$ are the pressures at the proximal and distal nodes of the pipe segment, respectively. Let $\mathcal{A}_i$ be the set of pipe segments emanating from the $ith$ pipe segment in the airway network. We can express the conservation of flow in the airway network as

$$Q_i = \sum_{j \in \mathcal{A}_i} Q_j. \tag{14}$$

The outlet pressure of the airway network is set using the boundary condition $P_0 = \hat{P}$.

#### 3.4.1 Coupling the airway network to the poroelastic model.

We introduce subdomains to identify the region of the domain that is supplied with fluid from a specific branch of the airway network and returns fluid through that branch. For notational purposes we use the subscript $di$ to indicate the most distal branches that have no further conducting branches emanating from them, but which enter a group of acinar units approximated by the continuous poroelastic model. We construct a Voronoi tesselation based on the $N$ terminal locations $\boldsymbol{y}_{di}, i = 1, \dots, N$ of the airway network. The $i$th subdomain $\Omega_i(t)$ is the subset of $\Omega$ that is closer the $i$th terminal location at $\boldsymbol{y}_{di}$ than to any of the other terminal locations, i.e,

$$\Omega_i(t) := \left\{ \boldsymbol{x} \in \Omega(t) : ||\boldsymbol{x} - \boldsymbol{y}_i|| < ||\boldsymbol{x} - \boldsymbol{y}_j||,\ j = 1, 2..., N\ , j \neq i \right\}, \quad i = 1, \dots, N. \tag{15}$$

Obviously we have $\overline{\Omega}(t) = \bigcup \overline{\Omega}_i(t)$. A simple 2D examples is shown in Figure 4.

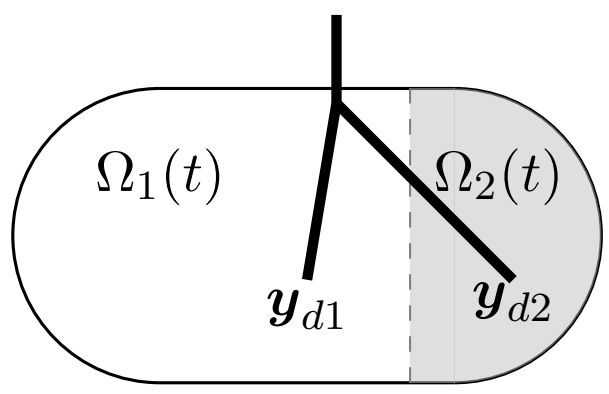

Figure 4: A simple example of a 2D domain being split into two subdomains according to (15).

We couple the airway network to the poroelastic domain in two ways. Firstly, the flux from each distal airway acts as a source term in the poroelastic mass conservation equation, namely

$$\nabla \cdot (\boldsymbol{\chi}_t + \boldsymbol{z}) = Q_{di} \quad \text{in } \Omega_i(t). \tag{16}$$

Secondly, the pressure at the distal airway $P_{di}$, determines the average pressure within subdomain $\Omega_i(t)$, i.e.,

$$\frac{1}{|\Omega_i(t)|} \int_{\Omega_i(t)} p \, \mathrm{d}\Omega_i(t) = P_{di}, \tag{17}$$

where $|\Omega_i(t)|$ denotes the volume of the subdomian $\Omega_i(t)$.

### 3.5 The coupled lung parenchyma / airway model

To solve the coupled poroelastic-fluid-network lung model we need to find $\boldsymbol{\chi}(\boldsymbol{X},t)$, $\boldsymbol{z}(\boldsymbol{x},t)$, $p(\boldsymbol{x},t)$, $P_i$ and $Q_i$ such that

$$\left.\begin{aligned}
-\nabla\cdot(\boldsymbol{\sigma}_e - p\boldsymbol{I}) &= \rho\boldsymbol{f} \quad \text{in } \Omega(t),\\
\boldsymbol{k}^{-1}\boldsymbol{z} + \nabla p &= \rho^f\boldsymbol{f} \quad \text{in } \Omega(t),\\
\nabla\cdot(\boldsymbol{\chi}_t + \boldsymbol{z}) &= Q_{di} \quad \text{in } \Omega_i(t),\\
\boldsymbol{\chi}(\boldsymbol{X},t)|_{\boldsymbol{X}=\boldsymbol{\chi}^{-1}(\boldsymbol{x},t)} &= \boldsymbol{X} + \boldsymbol{u}_D \quad \text{on } \Gamma_d(t),\\
(\boldsymbol{\sigma}_e - p\boldsymbol{I})\boldsymbol{n} &= \boldsymbol{t}_N \quad \text{on } \Gamma_n(t),\\
\boldsymbol{z}\cdot\boldsymbol{n} &= q_D \quad \text{on } \Gamma_f(t),\\
p &= p_D \quad \text{on } \Gamma_p(t),\\
\boldsymbol{\chi}(\boldsymbol{X},0) &= \boldsymbol{X}, \quad \text{in } \Omega(0),\\
P_0 &= \hat{P},\\
P_{i,1} - P_{i,2} &= R_i Q_i,\\
Q_i &= \sum_{j\in\mathcal{A}_i} Q_j,\\
\frac{1}{|\Omega_i(t)|}\int_{\Omega_i(t)} p \,\mathrm{d}\Omega_i(t) &= P_{di},
\end{aligned}\right\} \qquad (18)$$

where $\partial\Omega(t) = \Gamma_d(t)\cup\Gamma_n(t)$ for the mixture and $\partial\Omega(t) = \Gamma_p(t)\cup\Gamma_f(t)$ for the fluid, with an outward pointing unit normal $\boldsymbol{n}$.

## 4 Numerical solution of the poroelastic lung model

### 4.1 Discretization of the poroelastic equations

The poroelastic equations (8) are discretized, in space, using the finite element method. In particular, we use the stabilised mixed finite element $P1-P1-P0$. This element uses continuous piecewise linear approximations for both the fluid flux and deformation, $P1$. The pressure is approximated using the space of piecewise constants, $P0$. To provide a stable approximation scheme, stabilisation of the pressure is applied, for further details see [23]. The temporal discretisation is achieved by taking a standard backward Euler a approach.

The resulting system, including airway coupling and linearisation, is presented in sub-section .

### 4.2 Coupling of the fluid network to the discretized poroelastic model

As an example we discretize a 2D domain using triangular elements in Figures 5a and 5b. The $i$th discretized subdomain $\Omega_i(t)$ is defined as the set of elements $E_k$ whose centroids $\overline{\boldsymbol{x}_k}$ are closer to the distal end of the $i$th terminal branch than to the distal end of any other terminal branch, i.e.,

$$\Omega_i(t) := \left\{E_k \in \Omega(t) : ||\overline{\boldsymbol{x}_k} - \boldsymbol{y}_{di}|| < ||\overline{\boldsymbol{x}_k} - \boldsymbol{y}_{dj}||,\ j = 1,2...,N\,, j \neq i\right\}. \qquad (19)$$

The discretized subdomains converge towards the subdomains defined by (15) as the mesh is refined.

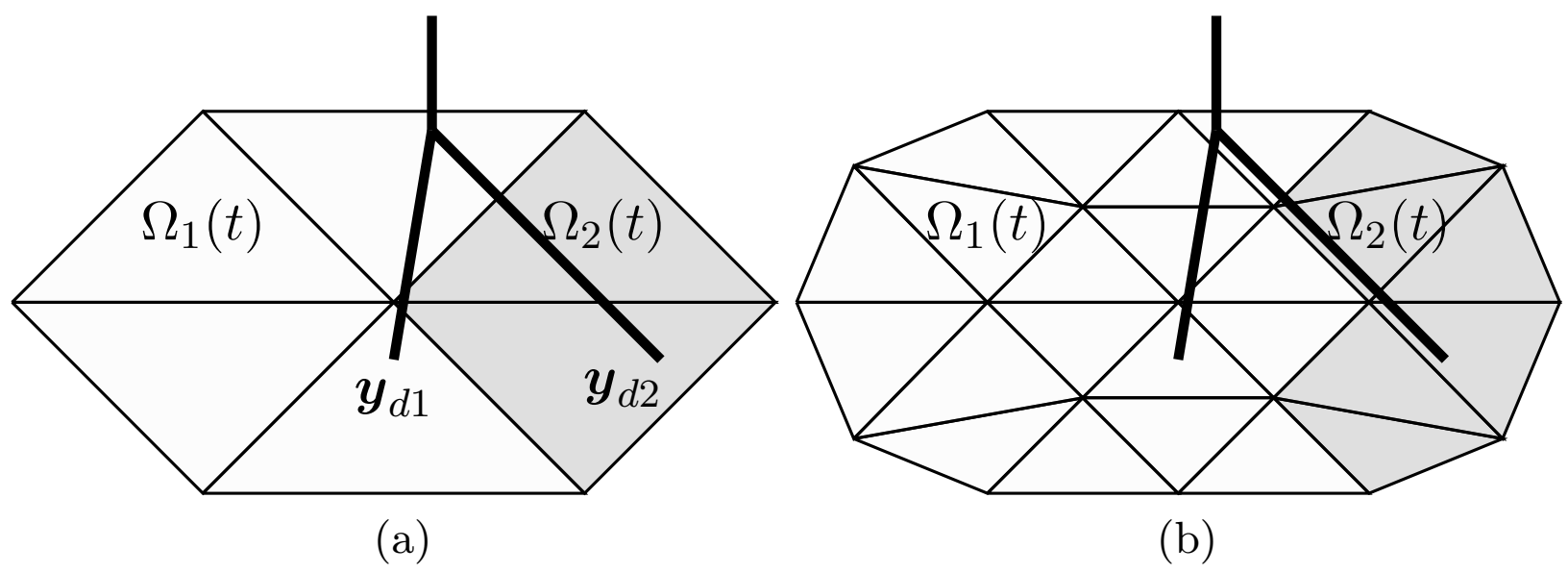

Figure 5: (a) Coupling between the discretized domain and the airway network for the example shown in Figure 4. (b) Coupling between the discretized domain and the airway network after mesh refinement.

### 4.3 Newton's method

The coupled nonlinear system of equations governing both deformation and flow in the poroelastic parenchyma, and flow in the airway tree at each $t^n$, is solved via Newton's method. Let $\mathfrak{u}_i^n$ denote the fully discrete solution of the coupled problem at the $i$th step within the Newton method at time $t^n$. The Newton algorithm at a particular time step $n$, is given in Algorithm 1.

---

**Algorithm 1** Newton algorithm at $t_n$

$i = 0$
$\mathfrak{u}_0^n = \{\boldsymbol{\chi}^{n-1}, \boldsymbol{z}^{n-1}, p^{n-1}\}$
**while** $||\boldsymbol{R}(\mathfrak{u}_i^n, \mathfrak{u}^{n-1})|| > \text{TOL}\ \&\ i < \text{ITEMAX}$ **do**
  Assemble $\boldsymbol{R}(\mathfrak{u}_i^n, \mathfrak{u}^{n-1})$ and $\boldsymbol{K}(\mathfrak{u}_i^n)$
  Solve $\boldsymbol{K}(\mathfrak{u}_i^n)\delta\mathfrak{u}_{i+1}^n = -\boldsymbol{R}(\mathfrak{u}_i^n, \mathfrak{u}^{n-1})$
  Compute $\mathfrak{u}_{i+1}^n = \mathfrak{u}_i^n + \delta\mathfrak{u}_{i+1}^n$
  Update the mesh, $\Omega(t_n) = \boldsymbol{X} + \boldsymbol{\chi}_i^n$
  $i = i + 1$
**end while**

---

At each Newton iteration we solve the linear system

$$\boldsymbol{K}(\mathfrak{u}_i^n)\delta\mathfrak{u}_{i+1}^n = -\boldsymbol{R}(\mathfrak{u}_i^n, \mathfrak{u}^{n-1}). \tag{20}$$

where $\boldsymbol{K}$ is the approximate linearization of the nonlinear system of equations, and $\boldsymbol{R}$ is the residual vector. In our implementation, the linearization is approximate since only the elasticity term is treated as a nonlinearity (see [23] for details).

### 4.4 The matrix system

Let $\boldsymbol{\phi}_k$ denote a vector-valued linear basis function, and

$$\boldsymbol{\chi}_i^n = \sum_{k=1}^{n_\chi} \boldsymbol{\chi}_{i,k}^n \boldsymbol{\phi}_k, \quad \boldsymbol{z}_i^n = \sum_{k=1}^{n_z} \boldsymbol{z}_{i,k}^n \boldsymbol{\phi}_k.$$

Similarly, let $\psi_i$ denote a piecewise constant basic function, so that

$$\boldsymbol{p}_i^n = \sum_{k=1}^{n_p} p_{i,k}^n \psi_k,$$

and let $\epsilon_k$ be a scalar-valued linear basis function so that

$$\boldsymbol{\Lambda}_i^n = \sum_{k=1}^{n_\Lambda} \Lambda_{i,k}^n \epsilon_k$$

is the Lagrange multiplier associated with the fluid flux boundary condition (see [23] for details). Let $\mathbf{P}^n$ and $\mathbf{Q}^n$ be the pressures at each junction and the fluid fluxes in each branch of airway network, except for the pressures at the distal end of, and the fluxes in, the most distal branches of the airway network which are given by $\mathbf{P}_d^n$ and $\mathbf{Q}_d^n$ respectively. The linear system (20) at each Newton iteration can be expanded as

$$\begin{bmatrix} \mathbf{K}^e & 0 & \mathbf{B}^T & 0 & 0 & 0 & 0 & 0 \\ 0 & \mathbf{M} & \mathbf{B}^T & \mathbf{L}^T & 0 & 0 & 0 & 0 \\ -\mathbf{B} & -\Delta t\mathbf{B} & \mathbf{J} & 0 & 0 & 0 & 0 & -\Delta t\mathbf{G}^T \\ 0 & \mathbf{L} & 0 & 0 & 0 & 0 & 0 & 0 \\ 0 & 0 & 0 & 0 & \mathbf{T}_{11} & \cdots & \cdots & \mathbf{T}_{14} \\ 0 & 0 & 0 & 0 & \vdots & & & \vdots \\ 0 & 0 & 0 & 0 & \mathbf{T}_{31} & \cdots & \cdots & \mathbf{T}_{34} \\ 0 & 0 & \mathbf{G} & 0 & 0 & -\mathbf{X} & 0 & 0 \end{bmatrix} \begin{bmatrix} \delta\chi^n \\ \delta\mathbf{z}^n \\ \delta\mathbf{p}^n \\ \delta\boldsymbol{\Lambda}^n \\ \delta\mathbf{P}^n \\ \delta\mathbf{P}_d^n \\ \delta\mathbf{Q}^n \\ \delta\mathbf{Q}_d^n \end{bmatrix} = - \begin{bmatrix} \boldsymbol{r}_1 \\ \boldsymbol{r}_2 \\ \boldsymbol{r}_3 - \Delta t\mathbf{G}^T\mathbf{Q}_d^n \\ 0 \\ 0 \\ 0 \\ 0 \\ \mathbf{G}\mathbf{p}^n - \mathbf{X}\mathbf{P}_d^n \end{bmatrix},$$

where

$$\begin{aligned}
\boldsymbol{k}^e_{kl} &= \int_{\Omega(t_n)} \left[\boldsymbol{E}^T_k \boldsymbol{D}(\boldsymbol{\chi}^n_i)\boldsymbol{E}_l + (\nabla\boldsymbol{\phi}_k)^T \boldsymbol{\sigma}_e(\boldsymbol{\chi}^n_i)\nabla\boldsymbol{\phi}_l\right] \, \mathrm{d}\Omega(t_n), \\
\boldsymbol{m}_{kl} &= \int_{\Omega(t_n)} \boldsymbol{k}^{-1}(\boldsymbol{\chi}^n_i)\boldsymbol{\phi}_k \cdot \boldsymbol{\phi}_l \, \mathrm{d}\Omega(t_n), \\
\boldsymbol{b}_{kl} &= -\int_{\Omega(t_n)} \psi_k \nabla \cdot \boldsymbol{\phi}_l \, \mathrm{d}\Omega(t_n), \\
\boldsymbol{j}_{kl} &= \Upsilon \sum_{K\in\mathcal{T}^h_i} \int_{\partial K\setminus\partial\Omega(t_n)_i} h_{\partial K} [\![\psi_k]\!][\![\psi_l]\!] \, \mathrm{d}s. \\
\boldsymbol{r}_{1i} &= \int_{\Omega(t_n)} \left[(\boldsymbol{\sigma}_e(\boldsymbol{\chi}^n_i) - p^n_i \boldsymbol{I}) : \nabla\boldsymbol{\phi}_i - \rho(\boldsymbol{\chi}^n_i)\boldsymbol{\phi}_i \cdot \boldsymbol{f}\right] \, \mathrm{d}\Omega(t_n) - \int_{\Gamma_n(t_n)} \boldsymbol{\phi}_i \cdot \boldsymbol{t}_N(\boldsymbol{\chi}^n_i) \, \mathrm{d}\Gamma_n(t_n), \\
\boldsymbol{r}_{2i} &= \int_{\Omega(t_n)} \left[\boldsymbol{k}^{-1}(\boldsymbol{\chi}^n_i)\boldsymbol{\phi}_i \cdot \boldsymbol{z}^n_i - p^n_i \nabla \cdot \boldsymbol{\phi}_i - \rho^f(\boldsymbol{\chi}^n_i)\boldsymbol{\phi}_i \cdot \boldsymbol{f}\right] \, \mathrm{d}\Omega(t_n), \\
\boldsymbol{r}_{3i} &= \int_{\Omega(t_n)} \psi_i \left[\nabla \cdot \left(\boldsymbol{\chi}^n_i - \boldsymbol{\chi}^{n-1}\right) + \Delta t \psi_i \nabla \cdot \boldsymbol{z}^n_i - \Delta t \psi_i g\right] \, \mathrm{d}\Omega(t_n) \\
&\qquad + \Upsilon \sum_{K\in\mathcal{T}^h} \int_{\partial K\setminus\partial\Omega(t_n)} h_{\partial K} [\![\psi_i]\!][\![p^n_i - p^{n-1}]\!] \, \mathrm{d}s, \\
\mathbf{l}_{ij} &= \int_{\Omega(t_n)} \epsilon_i \boldsymbol{\phi}_j \cdot \boldsymbol{n}, \, \mathrm{d}\Omega(t_n), \\
\mathbf{x}_{ij} &= \begin{cases} 1 & \text{if } ||\boldsymbol{y}_{di} - \overline{\boldsymbol{x}_j}|| < ||\boldsymbol{y}_{dk} - \overline{\boldsymbol{x}_j}||, k = 1, 2..., N \;, k \neq i, \\ 0 & \text{otherwise,} \end{cases} \\
\mathbf{g}_{ij} &= \int_{\Omega(t_n)} \mathbf{x}_{ij} \frac{\phi_j}{|E_j|} \, \mathrm{d}\Omega(t_n),
\end{aligned}$$

and $\mathbf{T}$ represents the matrix entries arising from equations (13) and (14). The matrices $\boldsymbol{D}$ and $\boldsymbol{E}$ are given in appendix A by equations (29) and (30), respectively. The matrix block $\boldsymbol{J}$, which makes use of the jump operator $[\![\cdot]\!]$, the stabilization parameter $\Upsilon$ and the diameter (3D) of the $K$th element $h_{\partial K}$, is required to stabilize the system, see [29] and [23] for details. Finally, $\overline{\boldsymbol{x}_j}$ denotes the centroid of the $j$th element.

# 5 Lung Structure and Modelling Parameters

## 5.1 Mesh generation

We developed the geometry of a whole organ model of the right lung from high-resolution CT images taken at total lung capacity (TLC) and at functional residual capacity (FRC). The lung was first segmented from the CT data (slice thickness and pixel size 0.73 mm) using the commercially available segmentation software Mimics[1], and we then used the open-source image processing toolbox iso2mesh [30] to generate a Tetrahedral mesh containing 38369 elements. The conducting airways were also segmented from the CT data taken at TLC, and a centerline with radial information was calculated. To approximate the remaining airways up to generation 8-13 we used a volume filling airway generation algorithm to generate a mesh of the airway tree containing 13696 nodes with 2140 terminal branches [31].

### 5.2 Reference state, initial conditions and boundary conditions

The poroelastic framework we have described requires a stress free reference state. Biological tissues do not possess a “reference state” where the material is free of both stress and strain, rather the cells that make up tissues are born into stressed states and live out their lives in these stressed states [32]. In order to define a stress-free reference state we scale the lung from FRC to a configuration in which the internal stresses and strains are assumed to be zero. The lung model is then uniformly inflated from the reference state to create a pre-stressed FRC configuration which has a mean elastic recoil of approximately $0.49 \times 10^3$ Pa, commonly understood to be a typical value [33]. From there we simulate tidal breathing. A similar approach has also been used in [8].

We register the expiratory (FRC) segmentation to the segmentation at TLC using a simple procedure that uses independent scalings in the $x, y$ and $z$ direction to map between the bounding boxes of the segmentations at FRC and TLC. This yields a rough estimate of the deformation field for the lung surface from expiration to inspiration. To simulate tidal breathing we assume a sinusoidal breathing cycle and expand the lung surface from FRC to 40% of the deformation from FRC to TLC. Specifically,

$$\boldsymbol{u}_D(t) = 0.2\left(1 + \sin(\frac{\pi}{2}(t+3))\right)\boldsymbol{u}_{D,TLC} \quad \text{on } \Gamma_d(t), \tag{21}$$

where $\boldsymbol{u}_{D,TLC}$ is the deformation of the lung surface from FRC to TLC. This results in a physiologically realistic tidal volume of 0.59 liters at a breathing frequency of 15 breaths per minute. We simulate breathing for a total of eight seconds or two breathing cycles. Due to the incompressibility of the poroelastic tissue, this also determines the total volume of air inspired/expired and the flowrate at the trachea, see Figure 7a and 7b respectively. We assume that no fluid escapes from the lung (except via the trachea) and impose zero flux boundary conditions at the lung surface. The outlet pressure of the airway network is set to zero (atmospheric pressure).

### 5.3 Simulation parameters

Several parameters for lung tissue elasticity and poroelasticity have been proposed [34, 5, 35, 10, 36], although there is no consensus in the values in the literature. In this study, we have chosen parameters as shown in Table 1 which result in physiologically realistic simulations.

| Parameter | Value | Reference |
|---|---|---|
| $\phi_0$ | 0.99 | [35] |
| $\kappa_0$ | $10^{-5}$ m$^3$ s kg$^{-1}$ | [35] |
| $E$ | $0.73 \times 10^3$ Pa | [36] |
| $\nu$ | 0.3 | [36] |
| $\mu_f$ | $1.92 \times 10^{-5}$ kg m$^{-1}$ s$^{-1}$ | [2] |
| $T$ | 8s | - |
| $\Delta t$ | 0.2s | - |
| $\Upsilon$ | $10^{-5}$ | - |

Table 1: Simulation parameters.

[1] http://biomedical.materialise.com/mimics

# 6 Model exploration

In the subsequent analysis the total and elastic stress is calculated as $\sqrt{\lambda_1^2 + \lambda_2^2 + \lambda_3^2}$, where $\lambda_1, \lambda_2, \lambda_3$ are the three eigenvalues of the stress tensor. We define the relative Jacobian $J_V = J/J_{FRC}$ which is the volume ratio between the current state and FRC, as a measure for ventilation. Running simulations over many breaths, we found that differences between the second breath and subsequent breaths were negligible, and we therefore present results from the second breath, $t = 4$s to $t = 8$s. Unless otherwise stated, all subsequent figures that do not explicitly state the time are taken at $t = 5.8$s, just before peak inhalation of the second time breath of the simulation. The sagital slices used in subsequent figures is shown in Figure 6a.

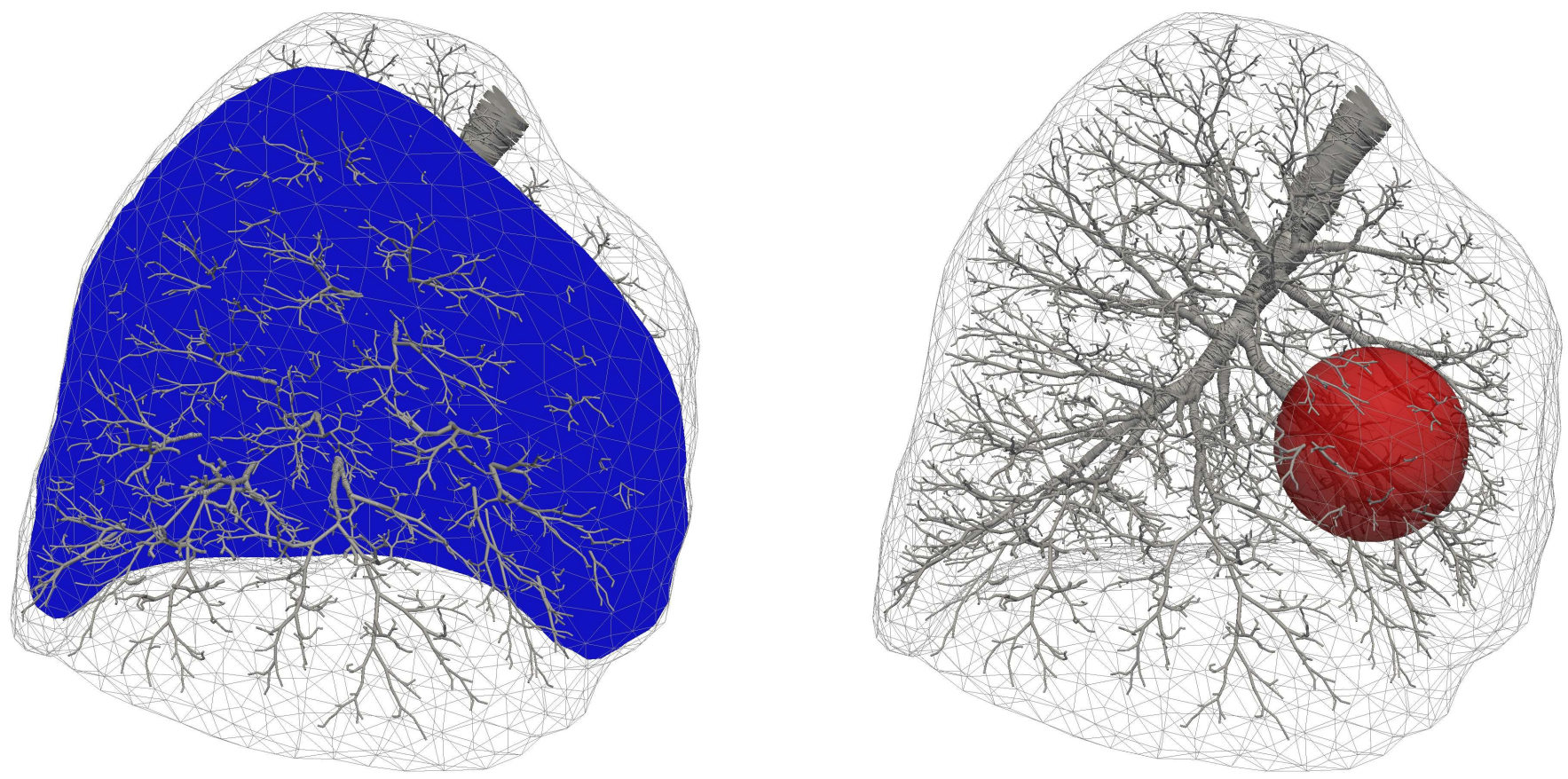

(a) (b)

Figure 6: (a) The blue sagital slice indicates the position of subsequent slices used for the data analysis of the tissue. (b) The red ball represents the structurally modified region used to prescribe airway constriction and tissue weakening.

## 6.1 Normal breathing

To simulate tidal breathing we apply the boundary conditions and simulation parameters previously discussed in sections 5.2 and 5.3, respectively.

Figure 7 details the lung tidal volume, flow rate and pressure drop obtained from simulations of tidal breathing. Due to the incompressibility of the poroelastic medium and the fixed nature of the airway network, the lung tidal volume (Figure 7a) and flow rate (Figure 7b) follow a sinusoidal pattern that matches the form of the deformation boundary condition prescribed by equation (21). The mean pressure drop in the airways is shown in Figure 7c, and agrees with previous simulation studies on full airway trees [1, 2].

The airway resistance (Poiseuille flow resistance) from the inlet (right bronchus) to each terminal airway is shown in Figure 8a for the entire airway. In Figure 8b we show the airway resistance of the terminal airways mapped onto the tissue.

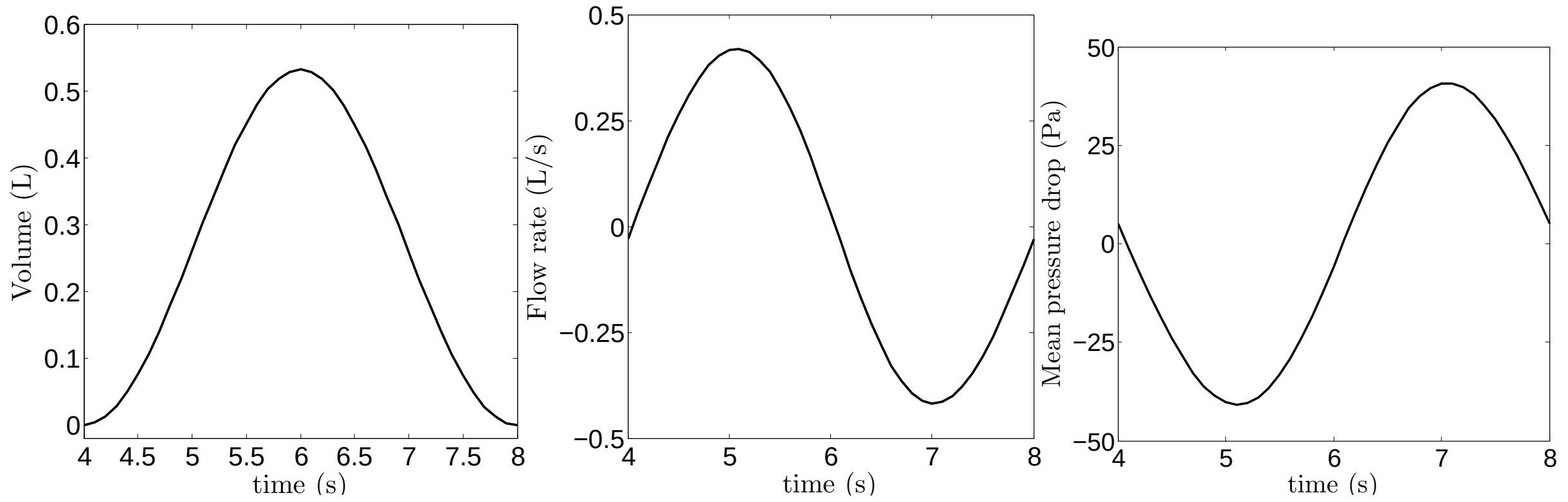

(a) (b) (c)

Figure 7: Simulated natural tidal breathing. (a) Lung tidal volume (volume increase from FRC). (b) Flow rate at the inlet. (c) Mean pressure drop from the inlet to the most distal branches.

In order to quantify the contribution of airway resistance to tissue expansion (ventilation), measured by $J_V$, the correlations between airway resistance in the tissue and $J_V$ are plotted for each element in Figure 9a. There is a clear correlation between airway resistance and tissue expansion, as is expected since the elastic coefficients are constant throughout the lung model. The Pearson correlation coefficients is $-0.55$, hence ventilation decreases as airway resistance increases, with a p-value $< 0.0001$. Figure 9b shows there is also a strong correlation between the airway resistance and pressure in the poroelastic tissue. Here the Pearson correlation coefficients is also $-0.55$, and pressure decreases (becomes more negative) with airway resistance, with a p-value $< 0.0001$. Note that for a very few regions that are coupled to terminal branches with a low airway resistance, positive pressures are possible. This results in a pressure gradient that pushes fluid from these well ventilated regions to neighbouring less ventilated regions, collateral ventilation, see [37].

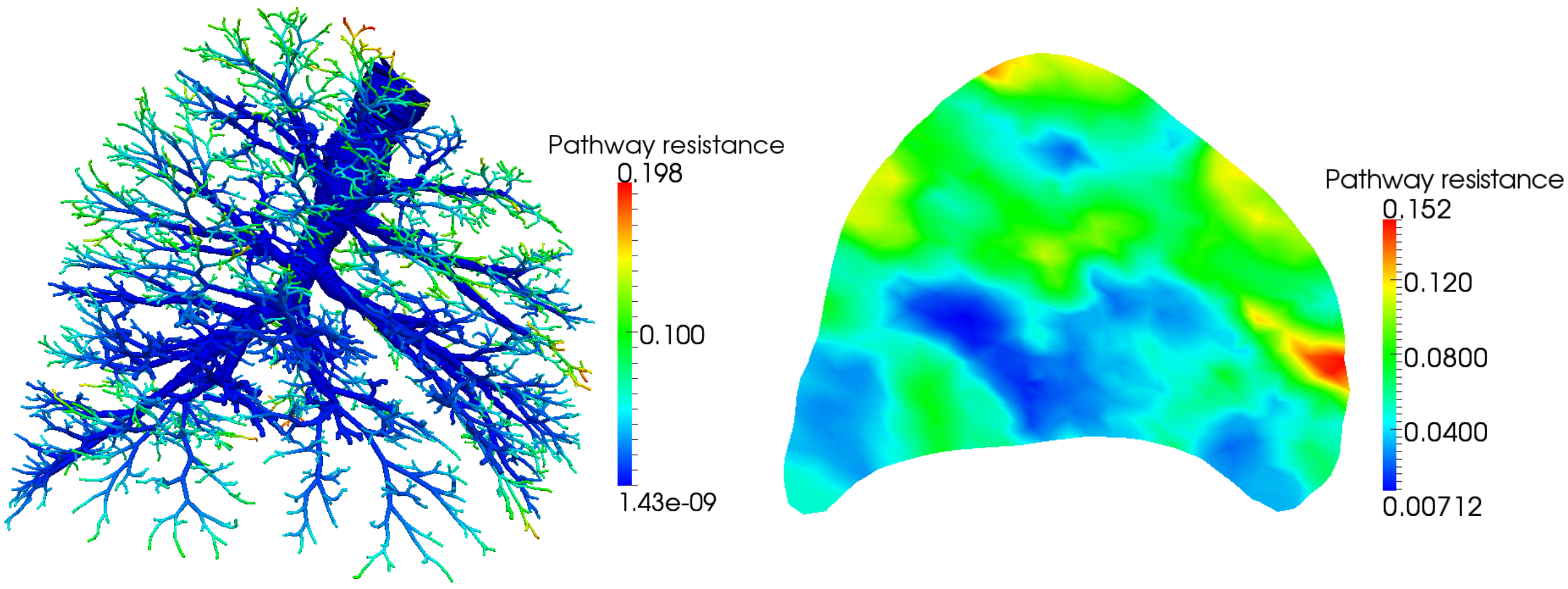

(a) (b)

Figure 8: (a) Pathway resistance ($\mathrm{Pa\,mm^{-3}s}$) from the inlet to the terminal branches in the airway tree. (b) Pathway resistance mapped onto a slice of tissue.

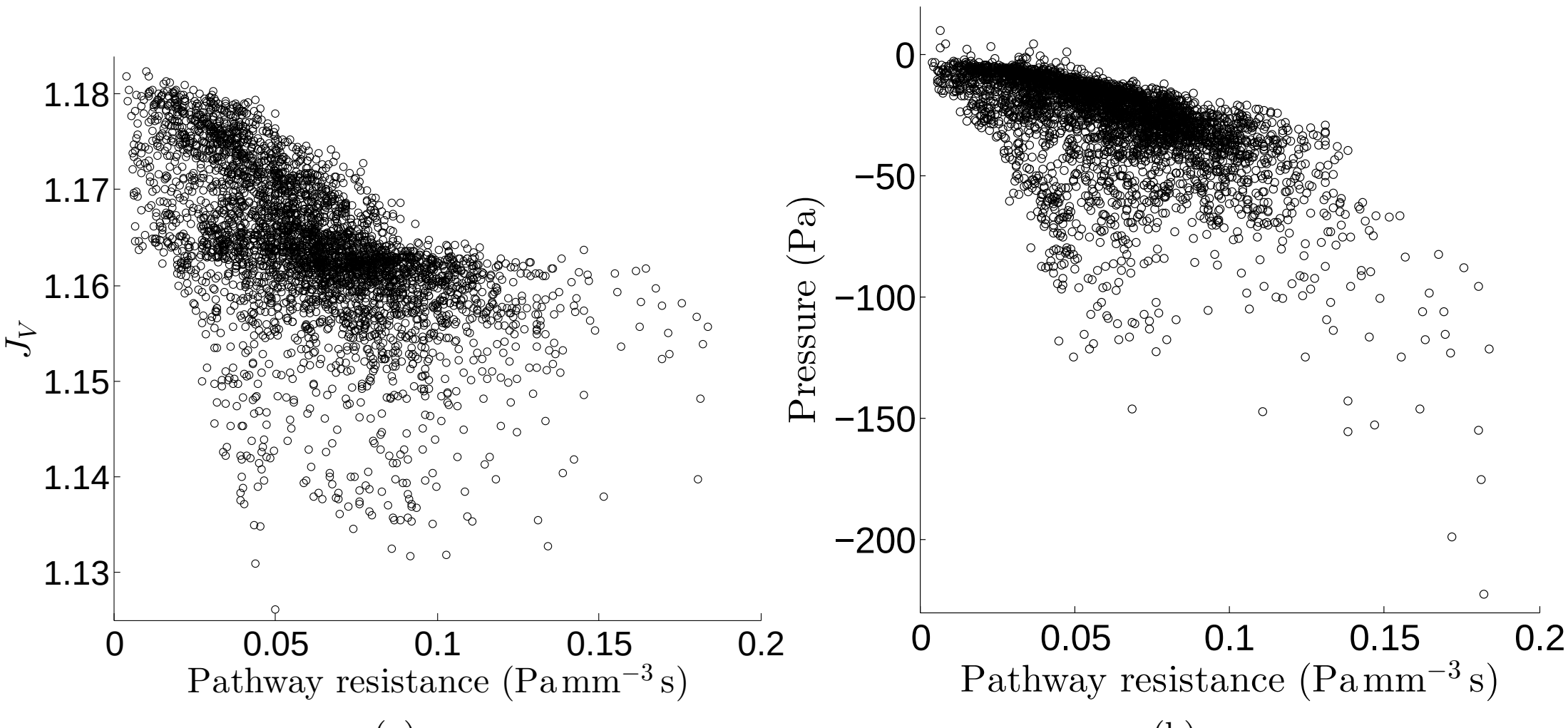

Figure 9: (a) Tissue expansion $J_V$ vs resistance of the pathways from the inlet to the terminal branch. (b) Pressure in the poroelastic medium (alveolar pressure) vs airway resistance.

The corresponding distribution of pressure in the airway tree is shown in Figure 10a and the pressure inside the poroelastic tissue is shown in Figure 10b. Figure 10c shows the pressure on the lung surface. The patchy pressure field is well approximated by the piecewise constant pressure elements employed by the finite element method used to solve the poroelastic equations (see [23]). Figure 10d shows the distribution of tissue expansion.

## 6.2 Breathing with airway constriction

We simulate localized constriction of the airways by reducing the radii of the lower airways, here airways with radius less than 4mm, within a ball near the right middle lobe. This region is represented by the red ball in Figure 6b. Reducing the radius increases airway resistance per unit length. We reduce the radius of the lower airways with radii less than 4mm by $0\%, 40\%, 50\%, 60\%$ and $65\%$. This corresponds to a mean airway resistance within the ball of $0.0507, 0.112, 0.188, 0.399$ and $0.651$ Pa mm$^{-3}$s, respectively. Mean tissue expansion during inspiration decreases as the airways become constricted as shown in Figure 11a. The standard deviation of local tissue expansion increases since the airway resistance of each branch increases by a different amount depending on its original length and radius. Long and narrow branches are affected most by the constriction. Since a larger pressure drop is needed to force the air down the constricted branches, the pressure decreases with increasing airway resistance as shown in Figure 11b. Figure 11c shows the elastic stress in the tissue decreases as airway resistance increases due to the decrease in tissue deformation (strain).

The simulations results shown in Figure 12 were performed with 65% airway constriction in the lower airways within the structurally modified region. The local volume conserving property (mass conservation) of the method is illustrated in Figure 12a where the tissue surrounding the constricted area is expanding to compensate for the reduction of tissue expansion due to the constriction within the structurally modified region. However, as seen in Figure 12c, a large elastic stress occurs near the boundary of the constricted region due to the difference in the relative expansions between the tissue within the modified region and the surrounding tissue. Figure 12b shows an increase in pressure near the boundary of this region. This facilitates air flow into the constricted region (collateral ventilation) to partially compensate for the reduced amount of ventilation, as is shown in Figure 12d. The magnitude of the maximum flow within the tissue is $8 \times 10^{-4}$ ms$^{-1}$ is quite small due to the low permeability of the poroelastic material.

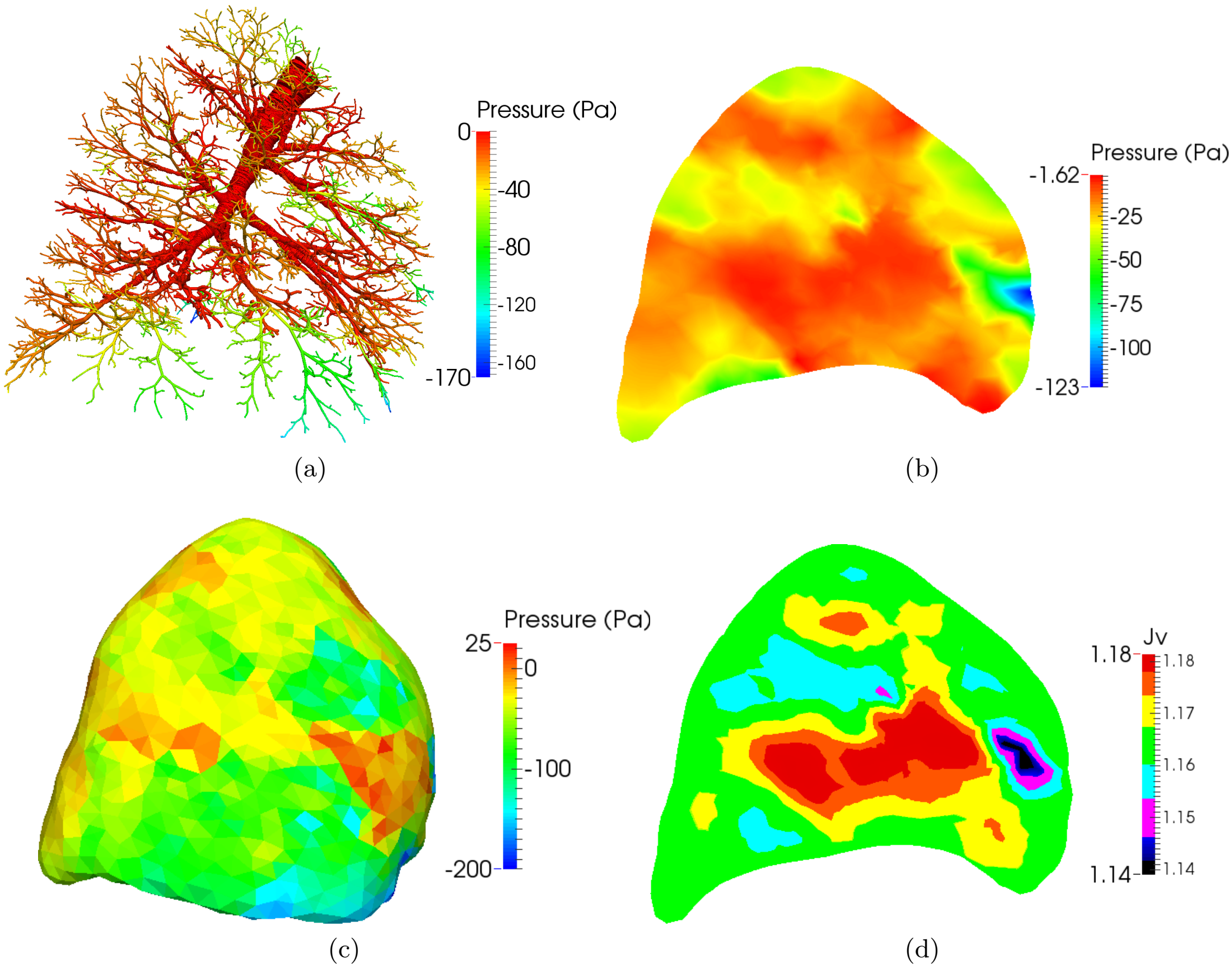

Figure 10: (a) Pressure in the airway tree. (b) Tissue pressure on the sagital slice. (c) Pressure on the lung surface. (d) Tissue expansion from FRC on the sagital slice.

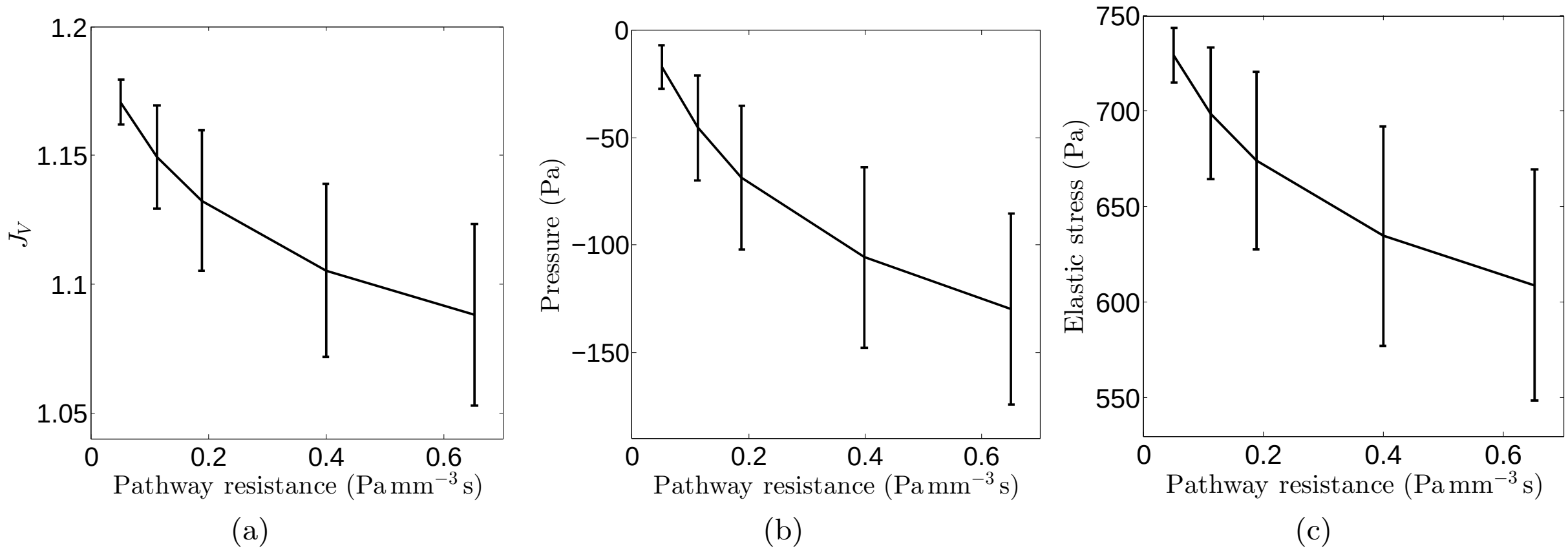

Figure 11: Global effects of increasing airway resistance within the modified region. (a) Means and standard deviations of tissue expansion, $J_v$. (b) Means and standard deviations of tissue pressures. (c) Means and standard deviations of elastic stresses.

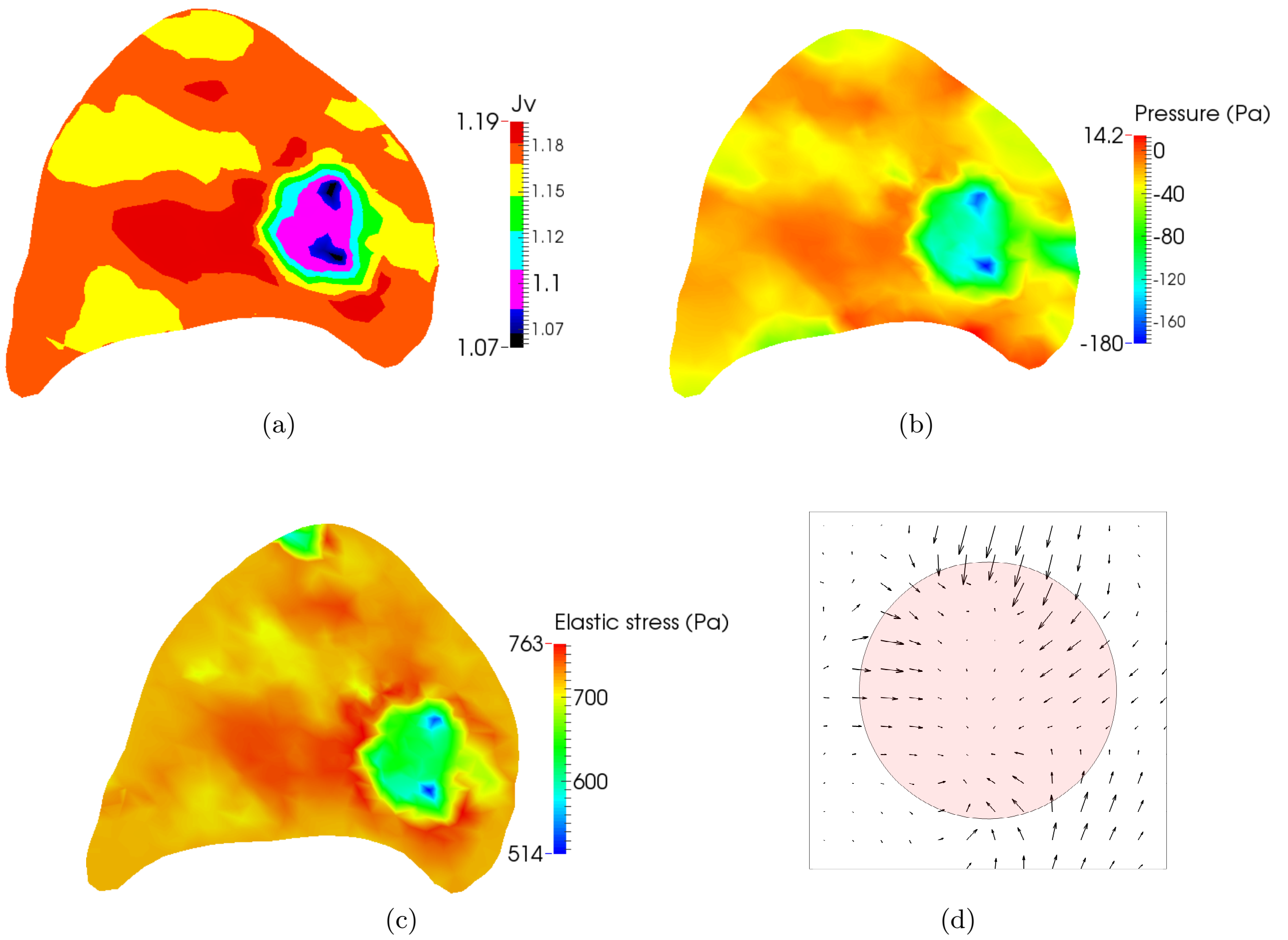

Figure 12: Effects in the sagital slice of a 65% airway constriction within the modified region.(a) Relative Jacobian $J_V$, (b) tissue pressure, (c) elastic stress, and (d) direction of the fluid flux near the structurally modified (constricted) region.

### 6.3 Breathing with locally reduced tissue elasticity

We simulate localized weakening of the tissue by reducing the Young's modulus of the tissue within the region represented by the red ball in Figure 6b. We reduce the Young's modulus by $0\%, 50\%, 75\%$ and $90\%$. This corresponds to a modified Young's modulus of $730, 365, 182.5$ and 73 Pa, respectively. Figures 13a-13c plot the local expansion $J_V$, the pressure and the elastic stress within the modified region. As expected, local expansion increases as the tissue weakens, and the elastic stress decreases. Note that in all cases the range (heterogeneity) of local ventilation, pressure and elastic stress within the modified region increases dramatically as the stiffness of the modified region decreases. In lung physiology such regions are said to be hyper-inflated. Due to the large amount of tissue expansion within the structurally modified region, the tissue immediately surrounding this region is effectively under-inflated between the expanded modified region and the surrounding tissue and as a result, as seen in Figure 14, it is this region which expands the least.

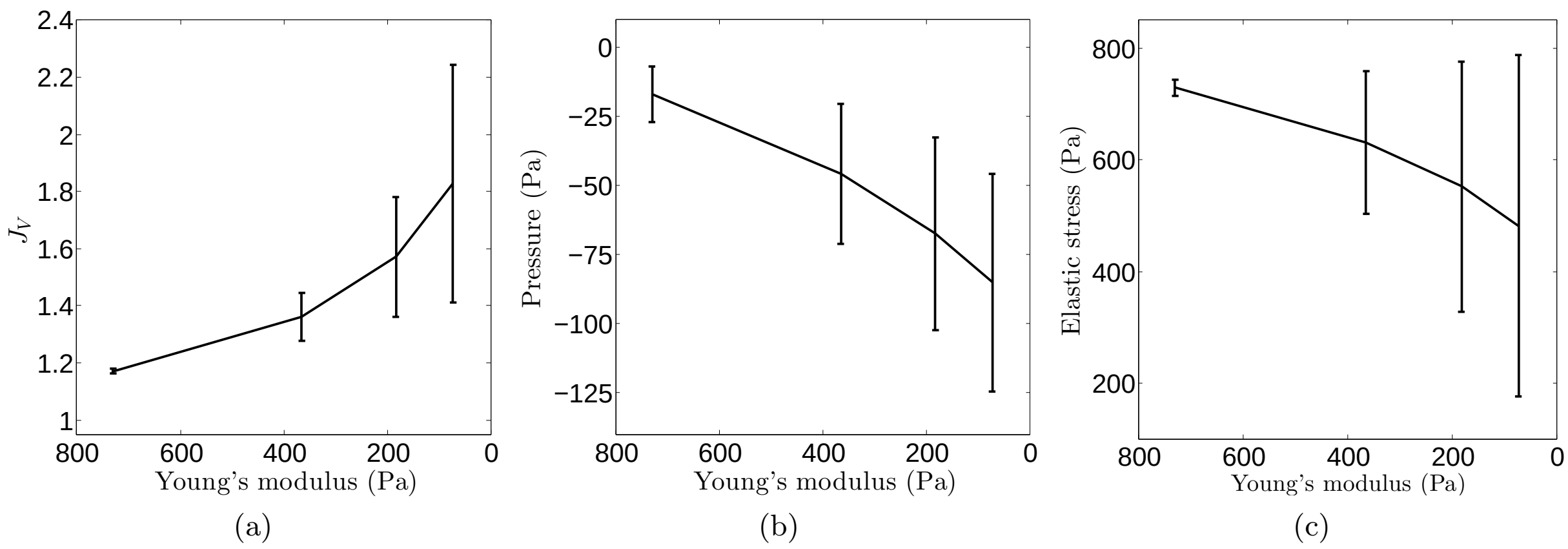

Figure 13: Effects of localized weakening of the tissue. Expansion, pressure and stress within the modified region vs Young's modulus within the structurally modified region. (a) Means and standard deviations of the relative Jacobian, (b) Means and standard deviations of the tissue pressure. (c) Means and standard deviations elastic stress.

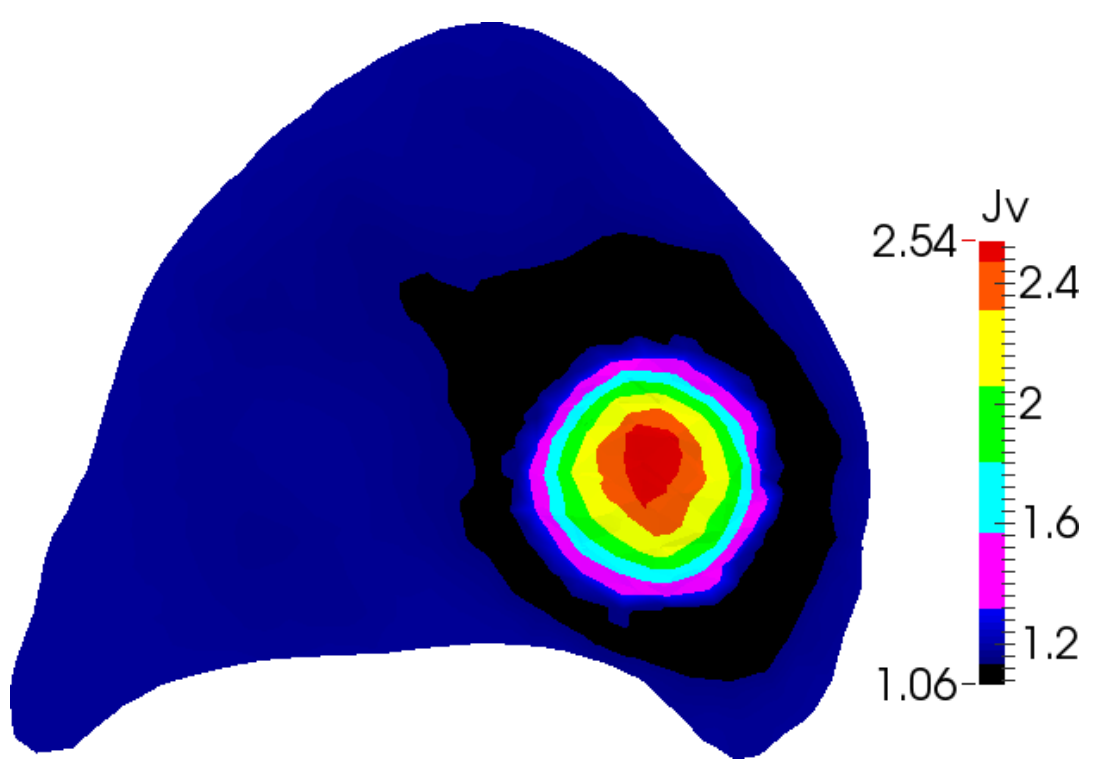

Figure 14: Relative Jacobian during inspiration with 90% localized tissue weakening.

### 6.4 Dynamic hysteresis

With the current choice of hyperelastic strain energy law (10) for the tissue mechanics, our model does not produce the classical hysteresis effects which are attributed to the visco-elastic properties of the biological tissues involved, elastin and collagen, within lung tissue [26]. However, we observe dynamic hysteresis effects caused by delayed emptying and filling of different regions of the lung.

Figure 15a plots the total stress against lung volume throughout the breathing cycle for three different breathing rates. The pressure-volume ($pV$) curves indicate very different amounts of hysteresis at different breathing rates.

Figure 15b and 15c both show the distribution of pressure against airway resistance shortly after the end of inhalation. At this point the lung as a whole has started to exhale air. However, particularly at the faster breathing rate, some lung regions have a negative pressure and are still filling. These parts of the lung also tend to have a higher airway resistance and therefore to fill more slowly and are subject to expansive stresses due to the more highly ventilated (and therefore expanded) tissue surrounding them. The negative pressures in the regions that are still filling near peak volume result in a larger total stress given by $\boldsymbol{\sigma} = \boldsymbol{\sigma}_e - p\boldsymbol{I}$, causing the curve to shift right when moving from inspiration to expiration (due to delayed filling) and left when moving from expiration to inspiration (due to delayed emptying). A faster breathing rate and therefore faster flow rates exaggerates this effect and causes an increased amount of hysteresis, i.e., a widening of the dynamic PV curve in Figure 15a. Since the work done is proportional to the area enclosed by the $pV$ diagram, considerably more work is required during more rapid breathing.

Figures 15b and 15c plot pressure against airway resistance for a four second and a one second breathing cycle respectively. We observe an increased heterogeneity in the tissue's pressure distribution with increased breathing rate. The pressure differences between poorly ventilated and well ventilated regions increase at high flow rates resulting in both an increase in stress (decrease in pressure) and an increase in heterogeneity of the total stress.

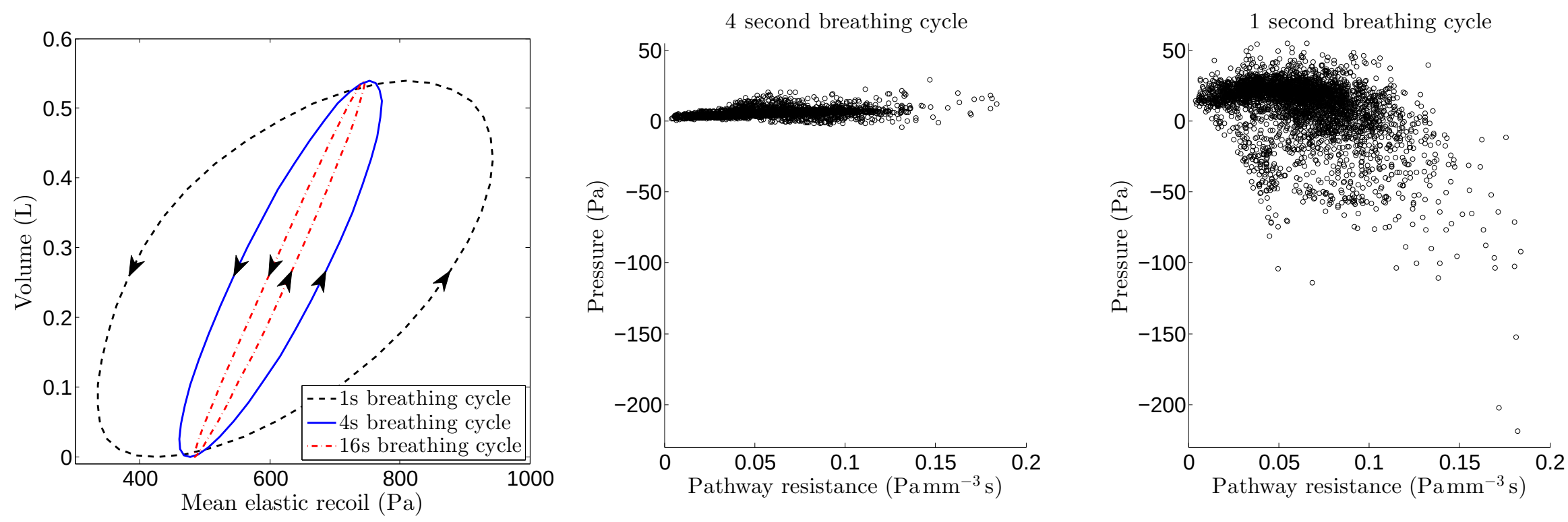

(a) (b) (c)

Figure 15: (a) Dynamic pressure-volume curve: mean elastic recoil (total stress) against lung tidal volume during one full breathing cycle, for three different breathing rates. The arrows indicate the direction of time during the breathing cycle. (b) Pathway resistance against pressure with a 4 second breathing cycle, 0.2 seconds after peak inhalation. (c) Pathway resistance against pressure with a 1 second breathing cycle, 0.05 seconds after peak inhalation.

# 7 Conclusions and Future Directions

We have presented a mathematical model of the lung that tightly couples tissue deformation with ventilation by using a poroelastic model for the lung parenchyma coupled to an airway network model for the bronchial tree. We have highlighted the assumptions necessary to arrive at such a model. In contrast with many previous ventilation models, the current approach models the tissue as a continuum. Therefore, it is able to locally conserve mass (which means conserve volume as the solid skeleton and fluid are both incompressible), and to model collateral ventilation. Further, it is driven by deformation boundary conditions extracted from imaging data to avoid having to prescribe a pleural pressure which is impractical to be measured experimentally. In simulations of normal breathing, the model is able to produce physiologically realistic global measurements and dynamics. In simulations with altered airway resistances and tissue stiffnesses, the model illustrates the interdependence of the tissue and airway mechanics and thus the importance of a fully coupled model. We have found that there is a strong correlation between airway resistance and the local mechanical properties of the tissue to ventilation, see Figure 9a.

Due to heterogeneity in airway resistance, dynamic hysteresis effects appear during breathing (Figure 15a) and result in a complex ventilation distribution caused by delayed filling and emptying of the tissue. In particular, small changes in airway radii result in large changes in pathway resistance, which in turn can significantly affect ventilation and mechanical stresses within the lung. Furthermore, we have shown that by reducing regional elastic properties of the tissue leads to areas of both hyper and under-inflation. Such physiological phenomenon are commonly seen within a diseased lung. Hence, by further refinement of the model it is hoped that it will provide insight into diseases such as asthma and chronic obstructive pulmonary disease.

Future directions will aim to provide a more detailed model and useable model. For example parametrizing the airways correctly is of great importance. However this is notoriously difficult since CT data is only available down to the 5-6th generation, and small errors and biases in the segmentation, that get propagated by the airway generation algorithm, can have large influences in determining the simulation results. It is hoped that the model presented here can form the basis for studies on the importance of airway and tissue heterogeneity on lung function, testing of mechanical hypothesis for the progression of disease, and investigations into phenomena such as hyperinflation, fibrosis and constriction.

# A Matrix notation

The fourth-order spatial tangent modulus tensor $\Theta_{ijkl}$ can be written as (in component form, see [38, section 5.3.2] and [39, section 6.6] )

$$\Theta_{ijkl} = \frac{1}{J} F_{iI} F_{jJ} F_{kK} F_{lL} \mathbf{C}_{IJKL}, \tag{22}$$

where $\mathbf{C}$ is the associated tangent modulus tensor in the reference configuration, given by

$$\mathbf{C}_{IJKL} = \frac{4\partial^2 W}{\partial C_{IJ} \partial C_{KL}} + pJ \frac{\partial C^{-1}_{IJ}}{\partial C_{KL}}. \tag{23}$$

For the numerical examples we have used the following Neo-Hookean strain-energy law

$$W(\boldsymbol{C}) = \frac{\mu}{2}(\mathrm{tr}(\boldsymbol{C}) - 3) + \frac{\Lambda}{4}(J^2 - 1) - (\mu + \frac{\Lambda}{2})\ln(J - 1 + \phi_0). \tag{24}$$

Thus, the resulting effective stress tensor is given by

$$\boldsymbol{\sigma}_e = \frac{\Lambda}{2}\left(J - \frac{1}{J-1+\phi_0}\right)\boldsymbol{I} + \mu\left(\frac{\boldsymbol{C}^T}{J} - \frac{\boldsymbol{I}}{J-1+\phi_0}\right), \tag{25}$$

and the spatial tangent modulus tensor is given as

$$\boldsymbol{\Theta} = \boldsymbol{\Theta}_e + p(\boldsymbol{I}\otimes\boldsymbol{I} - 2\mathcal{Z}), \tag{26}$$

where

$$\boldsymbol{\Theta}_e = \left[\Lambda J - 2\mu\left(\frac{1}{2(J-1+\phi_0)} - \frac{J}{2(J-1+\phi_0)^2}\right)\right]\boldsymbol{I}\otimes\boldsymbol{I} + \left[\frac{2\mu}{J-1+\phi_0} - \Lambda(J - \frac{1}{J-1+\phi_0})\right]\mathcal{B}, \tag{27}$$

and

$$\mathcal{B}_{ijkl} = \frac{1}{2}(\delta_{ik}\delta_{jl} + \delta_{il}\delta_{jk}), \quad \mathcal{Z}_{ijkl} = \delta_{ik}\delta_{jl}, \quad \boldsymbol{I}\otimes\boldsymbol{I} = \delta_{ij}\delta_{kl}. \tag{28}$$

See [38, chapter 5] and [25, chapter 3] for further details.

To simplify the implementation of the spatial tangent modulus we make use of matrix Voigt notation. The matrix form of $\beta$ is given by $\boldsymbol{D}$, which can be written as (see [38, section 7.4.2])

$$\boldsymbol{D} = \frac{1}{2}\begin{pmatrix} 2\boldsymbol{\Theta}_{1111} & 2\boldsymbol{\Theta}_{1122} & 2\boldsymbol{\Theta}_{1133} & \boldsymbol{\Theta}_{1112}+\boldsymbol{\Theta}_{1121} & \boldsymbol{\Theta}_{1113}+\boldsymbol{\Theta}_{1131} & \boldsymbol{\Theta}_{1123}+\boldsymbol{\Theta}_{1132} \\ & 2\boldsymbol{\Theta}_{2222} & 2\boldsymbol{\Theta}_{2233} & \boldsymbol{\Theta}_{2212}+\boldsymbol{\Theta}_{2221} & \boldsymbol{\Theta}_{2213}+\boldsymbol{\Theta}_{2231} & \boldsymbol{\Theta}_{2223}+\boldsymbol{\Theta}_{2232} \\ & & 2\boldsymbol{\Theta}_{3333} & \boldsymbol{\Theta}_{3312}+\boldsymbol{\Theta}_{3321} & \boldsymbol{\Theta}_{3313}+\boldsymbol{\Theta}_{3331} & \boldsymbol{\Theta}_{3323}+\boldsymbol{\Theta}_{3332} \\ & & & \boldsymbol{\Theta}_{1212}+\boldsymbol{\Theta}_{1221} & \boldsymbol{\Theta}_{1213}+\boldsymbol{\Theta}_{1231} & \boldsymbol{\Theta}_{1223}+\boldsymbol{\Theta}_{1232} \\ & \text{sym.} & & & \boldsymbol{\Theta}_{1313}+\boldsymbol{\Theta}_{1331} & \boldsymbol{\Theta}_{1323}+\boldsymbol{\Theta}_{1332} \\ & & & & & \boldsymbol{\Theta}_{2323}+\boldsymbol{\Theta}_{2332} \end{pmatrix}. \tag{29}$$

We also make use of the following implementation friendly matrix notation

$$\boldsymbol{E}_k = \begin{bmatrix} \phi_{k,1} & 0 & 0 \\ 0 & \phi_{k,2} & 0 \\ 0 & 0 & \phi_{k,3} \\ \phi_{k,2} & \phi_{k,1} & 0 \\ 0 & \phi_{k,3} & \phi_{k,2} \\ \phi_{k,3} & 0 & \phi_{k,1} \end{bmatrix}. \tag{30}$$